\documentclass[prl,twocolumn,aps,amssymb,footinbib]{revtex4}
\usepackage{amssymb}

\usepackage{graphicx}
\usepackage{amsmath}
\usepackage{times}

\begin{document}

\title{Quantum coherence of Hard-Core-Bosons and Fermions : Extended, Glassy and Mott Phases}

\author{ Ana Maria Rey$^{1,2}$, Indubala I. Satija$^{2,3}$ and Charles W. Clark$^{2}$} \affiliation{$^{1}$ Institute for
Theoretical Atomic, Molecular and Optical Physics, Cambridge, MA,
02138,USA} \affiliation{ $^{2}$ National Institute of Standards and
Technology, Gaithersburg MD, 20899, USA}
 \affiliation{$^{3}$ Dept. of Phys., George Mason U., Fairfax, VA, 22030,USA}

\date{\today}
\begin{abstract}

We use Hanbury-Brown-Twiss interferometry (HBTI) to study various
quantum phases of  hard core bosons (HCBs) and ideal fermions
confined in a one-dimensional quasi-periodic (QP) potential. For HCBs, the QP potential
induces a cascade of Mott-like band-insulator
phases in the extended regime, in addition to the Mott insulator, Bose glass, and superfluid phases.
At critical filling factors, the
appearance of these insulating phases is heralded by a peak to dip transition in the interferogram,
 which reflects the fermionic aspect of HCBs.   On the other hand, ideal fermions in the
extended phase display various complexities of incommensurate
structures such as devil's staircases and Arnold tongues. In the
localized phase, the HCB and the
fermion correlations are identical except for the sign of the peaks. Finally, we
demonstrate that HBTI provides an effective method to distinguish
Mott and glassy phases.

\end{abstract}
\pacs{71.30.+h, 03.75.Lm, 42.50.Lc}
\maketitle

Recent developments in the manipulations of ultra-cold atoms in
optical lattices have opened new possibilities for exploring the
richness and complexity of strongly-correlated systems. A particular
topic of continuous theoretical interest in condensed matter physics
has been the different phases  induced by the interplay between
disorder and interactions\cite{disorder,Lye}. In cold atomic gases
disorder can be introduced in a controlled manner either  by placing
a speckle pattern on the main lattice or by imposing a quasiperiodic
(QP) sinusoidal modulation on the original lattice  by using an
additional weaker lattice  with the desired wavelength ratio
\cite{Drese}. Furthermore, by changing the depth of the optical
potential it is experimentally possible to control the effective
dimensionality as well as the ratio between atomic kinetic and
interaction energy in these systems. All these attributes have  made
cold atomic systems a unique new laboratory to test many established
results, explore new phenomena underlying disordered systems and to
confront open questions\cite{Zoller,QPnew}.

In this paper, we use two and four point correlation functions to
compare and contrast the  different quantum phases in a gas of one
dimensional(1D) hard core bosons (HCBs) and fermions subject to a QP
potential. In an HCB gas \cite{GR}, the repulsive interactions
between bosons mimic the Pauli exclusion principle, and as a result
HCBs resemble in many respects a system of non-interacting fermions.
A 1D HCB gas in a periodic optical lattice has just recently been
experimentally realized \cite{Paredes}. A system of non-interacting
fermions might be realized experimentally by reducing interatomic
interactions via Feshbach resonances \cite{Fesh}. Here we treat the
case of atoms confined by a periodic lattice, with an additional QP
potential introduced to add pseudo-random disorder. A QP system is
in between periodic and random systems and is known to exhibit
localization transitions at a finite depth of the additional lattice
\cite{Harper}. Here we show that the interplay between the effects
of disorder, interactions  and quantum statistics leads to new
quantum phases, fractal structures such as the devil's staircase,
and fragmented Fermi distribution functions. Our studies reveal
these effects are manifested in first- and second-order
interferometry, which makes them accessible to direct experimental
observation. Furthermore, we find that QP disorder can be exploited
to distinguish Mott from glassy phases.

Four-point correlations can be experimentally probed by
Hanbury-Brown-Twiss interferometry (HBTI) \cite{HBT} which measures
the intrinsic quantum noise in intensity measurements, commonly
refereed to as {\it shot noise} or {\it noise correlations}. HBTI is
emerging as one of the most important tools to provide information
beyond that offered by standard momentum-distribution based
characterization of phase coherence. The latter is obtained from
images of the density distribution of the atomic sample after its
release from the confining potential. HBTI is performed by measuring
the density-density correlations in such images
\cite{Altman,Foelling}.  Sharp peaks (dips) in these correlations
reflect bunching (anti-bunching) of bosons (fermions), characterize
the underlying statistics, and reveal information about the spatial
order in the lattice \cite{Altman,reynoise}.

In our study  we find that the QP potential induces in the HBTI
pattern an hierarchical set of peaks appearing at the reciprocal
lattice vectors of both lattices with competing periodicities. We
use these  peaks together with the first order Bragg peaks in the
momentum distribution  to provide a detailed characterization of the
different  many-body phases. Three characteristic phases are known
in disordered HCB systems \cite{Fisher}: i) the incompressible Mott
Insulator (MI) phase, ii) the superfluid phase, and iii) the
insulating but compressible Bose-glass (BG) phase. Here we also find
a number of {\it Mott-like band insulator} phases. These phases
correspond to the filling of the sub-bands of a fragmented energy
spectrum that emerges from the QP superfluid phase. Transitions to
these phases are signaled by a decrease in the intensities of the
first and second order coherent peaks. They are found to be followed
by a peak to dip transition in the HBTI pattern when an empty band
is populated with few atoms.

Ideal fermions in the localized MI and glassy phases exhibit first- and second-order
interference patterns
similar to the ones of HCBs, except
for a different sign of the peaks. However, in the extended phase,
fermions and HCBs behave quite differently, with fermion interference
patterns displaying various complexities of incommensurate structures. When
plotted as a function of the filling factor, their quasi-momentum distribution
displays an Arnold
tongue-like structure; the intensity of HBTI peaks exhibits a
step-like pattern which resembles a  devil's staircase \cite{Ott}.
Thus, our study points out the potential of ultra-cold gases to
provide laboratory demonstrations of non-linear and multi-fractal
phenomena.

In a typical experiment, atoms are released by turning off the
external potentials. The atomic cloud expands, and is photographed
after it enters the ballistic regime. Assuming that the atoms are
noninteracting from the time of release, properties of the initial
state can be inferred from the spatial images
\cite{Altman,Roth,reynoise}: the column density distribution image
reflects the initial quasi-momentum distribution, $n(Q)$, and the
density fluctuations, namely the {\it noise correlations}, reflect
the quasi-momentum fluctuations,  $\Delta(Q,Q')$,

\begin{eqnarray}
\hat{n}(Q)&=&\frac{1}{L}\sum_{j,k} e^{i Q a(j-k)
}\hat{\Psi}_j^\dagger \hat{\Psi}_k,\\
 \Delta(Q,Q')&\equiv&
\langle\hat{n}(Q) \hat{n}(Q')\rangle
-\langle\hat{n}(Q)\rangle\langle\hat{n}(Q')\rangle. \label{nnoise}
\end{eqnarray}In Eq. (\ref{nnoise}) we have assumed that  both, $Q,Q'$ lie inside the first Brillouin
zone. In this paper we focus on the quantity $ \Delta(Q,0)\equiv
\Delta(Q)$. $L$  is the number of lattice sites and $a$ the lattice
constant. $\hat{\Psi}_j$ is a bosonic or  fermionic annihilator
operator at the site $j$. The low energy physics of a 1D gas of
strongly correlated bosons in an optical lattice modulated by a QP
potential is well described by the Hamiltonian:
\begin{equation}
H^{(HCB)}=-J\sum_j(\hat{b}_j^{\dagger}\hat{b}_{j+1}
+\hat{b}_{j+1}^{\dagger}\hat{b}_{j})+\sum_{j} V_j
\hat{n}_j,\label{HCB}
\end{equation}
where $\hat{b}_j$ is the annihilation operator at the lattice  site
$j$; it satisfies bosonic commutation relations, plus the on-site
condition ${\hat{b}_j}^2={\hat{b}_j^{\dagger}}{}^2=0$ which
suppresses multiple occupancies. Here $J$ is the  hopping energy
between adjacent sites, and we have introduced $V_j=2 \lambda \cos( 2\pi \sigma j)$
to describe the additional QP potential. The
parameter $\lambda$ is proportional to the intensity of the lasers
used to create the QP lattice \cite{Drese}. Note that the primary
optical lattice defines the tight-binding condition, and the
QP potential is created by a secondary optical lattice.  Ideal
fermions are described by the same Hamiltonian, in which the HCB operators
are replaced by fermion operators.

For a single atom, the eigenfunctions at site $j$,  $\psi_j^{(n)}$  and
eigenenergies $E^{(n)}$  of  the Hamiltonian in  Eq.(\ref{HCB})
satisfy:
\begin{equation}
-J(\psi_{j+1}^{(n)}+\psi_{j-1}^{(n)})+ 2 \lambda \cos(2\pi\sigma j )
\psi_{j}^{(n)}=E^{(n)} \psi_{j}^{(n)}.
\end{equation}This is the Harper equation, a paradigm in  1D
QP systems \cite{Harper}. For irrational $\sigma$, the model
exhibits a transition from extended to localized states at
$\lambda_c=J$. Below criticality,
 all states are extended Bloch-like states characteristic of a periodic potential.
Above criticality,  all states are exponentially localized and  the
spectrum is point-like. At criticality the spectrum is a Cantor set,
and the gaps form a devil's staircase.

In this letter we treat the case of $\sigma=(\sqrt{5}-1)/2$. In numerical
studies, $\sigma $ is approximated by the ratio of two Fibonacci
numbers $F_{M-1}/F_{M}$, ($F_1=F_0=1, F_{i+1}=F_{i}+F_{i-1}$), which
describe the best rational approximant to $\sigma$. For this rational approximation the unit cell
has length $F_M$ and the single-particle spectrum consists of $F_M$
bands and $F_M-1$ gaps. The gaps occur at half-values of reciprocal
lattice vectors, which we label by the Bloch index $Q_n= \pm n\sigma$(mod 1)$2\pi/a$,
with $n$ an integer and $Q_na \in (-\pi,\pi)$. The
size of the gaps decreases as $n$ increases. In this paper, we present our results
for $\sigma=55/89$.

In general  the phase diagram of interacting bosons with disorder is
rather complicated. However the analysis simplifies in the HCB
limit, because multiple site occupancies are forbidden and the
system is thus "fermionized". The ground state energy of an
$N$-particle HCB system is the sum of the first $N$ single-particle
eigenstates, as is the case for ideal fermions. This is why the
localization transition occurs at $\lambda_c=J$ in HCB systems.
Two-point and four-point correlation functions for HCBs can be
expressed as a T\"{o}plitz-like determinant involving two-point free
fermionic propagators: $ g_{lm}=\sum_{n=0}^{N-1} \psi_l^{*(n)}
\psi_m^{(n)} $. However, in contrast to the ideal fermion case where
a direct application of Wick's theorem can be used  to express any
 correlation function in terms of $g_{lm}$, for HCBs the
calculations are more elaborate because they require the evaluation
of multiple determinants \cite{LM,reynoise}.

\begin{figure}[htbp]
\includegraphics[width =0.7\linewidth,height=0.7\linewidth]{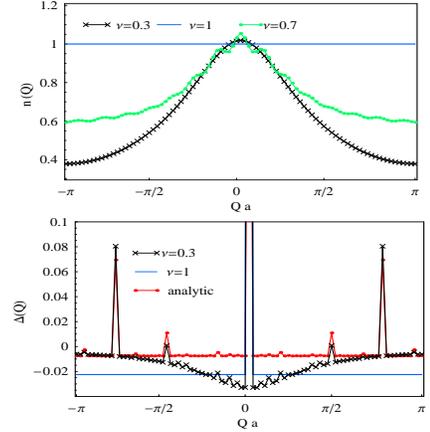}
\leavevmode \caption{(color online) Correlations of HCBs in the
glassy phase ($\lambda=2J$) Top: Momentum distribution. Bottom:
noise correlations. The central peak at $Q=0$ is truncated to
highlight the QP peaks.} \label{fig1}
\end{figure}

We now discuss the phase diagrams for HCB and fermion systemes, starting with the localized phase.
In the limit  $\lambda \gg \lambda_c$ , the single
particle wave functions are localized at  individual lattice
sites, $ \psi_j ^{(n)}\to \delta (j-l_n)$ and $E^{(n)} \to
2\lambda\cos(2\pi\sigma l_n)$, where  the localization center $l_n$
is such that $\cos(2\pi\sigma l_{n-1})<\cos(2\pi\sigma l_n)$. In
this limit the determinants involved in the evaluation of HCB
correlations  become trivial and an  analytic description is
possible. The  momentum distribution interference pattern flattens
out for both HCBs and fermions, $\hat{n}(Q)=\nu$ with $\nu=N/L$ the
filling factor. The  noise correlations simplify to $
\Delta(Q)=\nu \delta_{Q,0}-\frac{2\nu
(1-s)}{L}+\frac{(-1)^{s}}{L^2}\left |\sum_{l} e^{iaQl}
g_{ll}\right|^2$, where $s=0$ for HCBs and $s=1$ for fermions.
Approximating the sum by an integral, it is possible to show that
$\Delta(Q)$ is nonzero only at the reciprocal lattice vectors, $Q_n$, of the
combined superlattice. At these
points,
\begin{equation}
\Delta(Q_n)\equiv \Delta_n \approx
\nu\delta_{n,0}-\frac{2\nu(1-s)}{L}+ \frac{(-1)^{s}\sin^2[\pi \nu
n]}{(\pi n)^2}. \label{loc}\end{equation} Eq. (\ref{loc})
illuminates various important aspects of HBTI
 in the glassy phase: i) the explicit dependence of
noise correlations on quantum statistics, since the
boson peaks are positive and the corresponding fermion peaks
are negative (except for the autocorrelation peak $\Delta_0$);
ii) the potential of HBTI as an experimental
spectroscopy tool, since the intensity of the $n=1$ peak
 is directly related to the ground state energy of the many-body
system, $\Delta_{1}=\sum_{k=0}^{N-1} \cos(2\pi n \sigma l_k) =E/(L 2
\lambda)$; iii) the capability of noise-correlations to clearly
distinguish between  BG and MI phases, since in the Mott phase,
$\nu=1$, the whole hierarchical set of QP peaks disappears; iv) the
universal properties of the peaks whose intensities approach an
asymptotic value independently of the underlying commutation
relations (except for the $\Delta_0$ peak which is always larger for
HCBs). This is a
 unique feature of the BG phase, as in the extended phase fermionic dips
are weaker than the corresponding bosonic peaks (see below).

Fig.1 shows the momentum distribution, $n(Q)$  and noise
correlations,  $\Delta(Q)$, in the localized phase. Both $n(Q)$ and
$\Delta(Q)$ were computed numerically for finite $\lambda$ using the
theoretical framework discussed in Refs. \cite{reynoise,LM}. It
should be noted that even though the analytic results were derived
for $\lambda\to \infty$ limit, they provide a fair description of
the numerical results for finite $\lambda$. The main
difference between the $\lambda=2J$ HBTI pattern and the analytic
solution is the negative background which is larger at low
quasi-momenta. The background reflects the fact that the single
particle localized wave functions have finite localization length
$\xi =1/\log(\lambda)$. This also gives rise to the
finite width in the momentum distribution, which is proportional to $\xi$.
As shown in the figure, QP-induced localization may  lead to
Friedel oscillations \cite{Friedel} in the momentum distribution.
These oscillations are due to the two-fold degeneracy of the
eigenstates and can be understood in the strongly disordered limit
by taking the eigenfunctions to be a superposition of the two
degenerate wave functions: one localized at $l_k$ and the other at
$L-l_k$. For an odd number of atoms, this leads to $n(Q)=1+
 \frac{1}{L} \cos(2Q a l_{N-1} )$.

\begin{figure}[htbp]
\includegraphics[width =0.7\linewidth,height=0.7\linewidth] {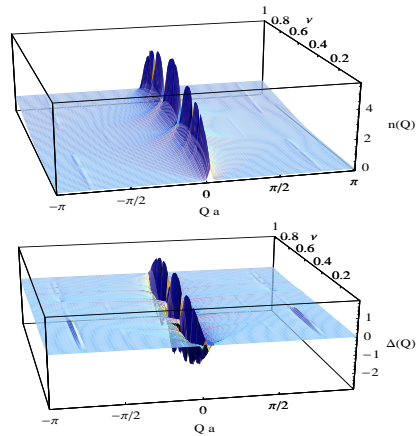}
\leavevmode \caption{(color online)HCBs in the superfluid phase,
$\lambda=0.5J$, top: Momentum distribution and bottom: noise
correlations where the intensities of the central peaks are scaled
by a factor of $1/10$ to show the satellite dips.
 } \label{fig2}
\end{figure}

\begin{figure}[htbp]
\includegraphics[width =0.8\linewidth]{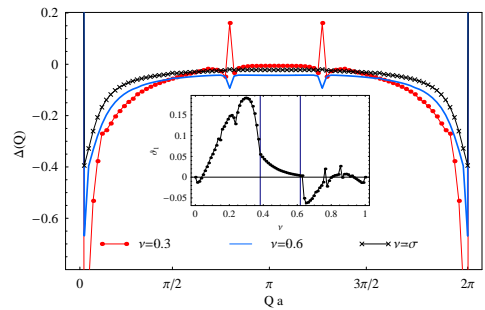}
\leavevmode \caption{(color online) HCBs noise correlation for
$\lambda=0.5J$.
 In the inset we plot the visibility for the dominant peak (see text)  as a function of $\nu$.} \label{fig3}
\end{figure}

In the extended phase two-point fermion correlations $g_{lm}$ are
long-range, the T\"oplitz-like determinants are complicated, and we
have found an analytical analysis to be difficult. In this phase our
numerically obtained  results  are summarized in Figs. 2 and 3. In
contrast to the glassy phase, the momentum distribution exhibits
interference peaks which signal the quasi-long-range coherence
characteristic of this phase:  a large peak at $Q=0$ and
quasiperiodicity induced peaks at reciprocal lattice vectors $\pm
\hbar Q_n$  whose intensity depends on the filling factor. These
peaks also exist in the noise interference pattern where they are
narrower and are accompanied by satellite dips, immersed in a
negative background. The later reflects the long range coherence in
the extended phase as discussed in Ref.\cite{reynoise}. It should be
noted that, in contrast to glassy phase, few QP related peaks are
visible.

A striking aspect of this phase is the cascade of lobes (seen in the
Fig.~2) describing a series of transitions to insulating phases.
These transitions occur at filling factors $\nu_c^{(n)}=n\sigma$(mod
1) and $\bar{\nu}_c^{(n)}=1-n\sigma$(mod 1), which correspond to the
fillings at which the various sub-bands underlying the QP spectrum
are completely filled. At these critical fillings, both first and
second order correlations depict a reduction in the central peak and
dip intensities and also the disappearance of the QP induced peaks
as shown in Fig.~2. Explicit computation of the compressibility
shows that the system is in fact incompressible at these critical
fillings. These features are reminiscent of the Mott insulating
phase at $\nu=1$, so we refer to these phases as {\it Mott-like
band-insulator} phases. However, in contrast to the Mott phase,
number fluctuations in these phases do not completely vanish, but
are only somewhat reduced.

A rather interesting consequence of the Mott-like band-insulating
transitions is the fact that as $\nu$  is increased beyond the
critical value, the peaks at the reciprocal lattice vectors
associated with the corresponding filled band become dips. This
change in the sign of the peaks beyond the critical filling can be
seen by plotting the fringe visibility $\vartheta_n=\Delta(Q_n)-
[\Delta(Q_n+\delta Q)- \Delta(Q_n-\delta Q)]/2$, with $\delta
Q=4\pi/aL$. Fig.~3 illustrates this for the dominant peaks that
occur at $\pm Q_{1}$. Here the peak to dip transition at
$\nu_c^{(1)}$, is signaled by a jump in the visibility from positive
to negative. This dip in the second order interference pattern can
be interpreted as a manifestation of fermionization. Physical
insight into this behavior can be gained by looking at the noise
correlation pattern of a single particle which  can be shown to
exhibit negative fermion-type dips: $\Delta(Q)=|z_0|^2 \delta_{Q0}
-|z_Qz_0|^2$ with $z_Q$ is the Fourier-transform of the ground state
wave function. For HCBs these negative fringes survive even for two
atoms.  This implies that the peak to dip transition can be
understood  as the consequence of an empty band populated with few
atoms.

In contrast to the   superfluid behavior of HCBs,  fermions in the
extended phase  have  metallic properties and  different
interferometric pattern. For $\lambda=0$, the sharply peaked HCB
momentum  distribution is replaced by a step-function profile for
fermions: $n(Q)=1$ for $|Q|\leq Q_F$ and $n(Q)=0$ for $|Q|> Q_F$,
with $ Q_F=\pi \nu/a$ being the Fermi momentum. For
$0<\lambda<\lambda_c$ this distribution retains part of the
step-like profile of the free fermion gas. However, the
 Fermi sea gets fragmented and additional step-function
structures centered at different reciprocal lattice vectors of the
second lattice develop (See Fig.4).  We call the filled states
centered around $Q=0$ the main Fermi sea and those around the QP
related reciprocal lattice vectors  the quasi-Fermi seas.

The quasi-Fermi seas emanating from the QP reciprocal lattice
vectors form tongue-like structures  which resemble in many respects
the Arnold tongues that describe mode-locked periodic windows in the
circle map \cite{Ott}. As the number of atoms increases, the width
of the tongues increase. Precisely  at the critical fillings
$\nu_c^{(n)}$ and $\bar{\nu}_c^{(n)}$, at which a quasi-band is
filled up and  the system becomes a band-insulator, a quasi-Fermi
sea merges with the main Fermi sea. As $\lambda$ increases, more
tongues become visible, and with the increase of $\nu$ they overlap
in a complex pattern. As seen in Fig.4 , we have tongues
corresponding to both the particles and holes, which together
describe the fragmented Fermi-Dirac distribution when both $Q$ and
$\nu$ vary.

Noise correlations for fermions ( Fig. 4) also bear a striking
contrast to those of HCBs ( Fig. 2 and 3). One sees a series of
plateaus at reciprocal lattice vectors, as $\nu$ is varied. The
origin of this step-like structure, where jumps occur at  the
filling factors $\nu_{ju}^{(n)}= |Q_n| a/\pi$, can be understood
from a perturbative argument as follows. For non-interacting
fermions, for $Q \ne 0$, $\Delta(Q,0)=-\sum_{k=0}^{N-1}|z_Q^{(k)}
z_0^{(k)}|^2$ with $z_Q^{(k)}$ being the Fourier transform of the
$k^{th}$ single-particle eigenfunction. For $\lambda=0$ ,  this
overlap between any  two different Fourier components  is always
zero as only the ground ($k=0$) state has a zero quasi-momentum
component, ie, $z_0^{(0)}=1$. For small $\lambda << J$, first order
perturbation theory explains the single step observed at
$\nu_{ju}^{(1)}=|Q_1| a/\pi$, as only $z_0^{(k_1)}$ with
$k_1=\nu_{ju}^{(1)} L$ is nonzero.  As $\lambda$ increases further,
more steps are observed as other eigenvectors, acquire  a nonzero
$Q=0$ component. The steps can also be seen in the $Q=0$ plane of
$n(Q)$. At criticality, the whole hierarchy of steps resembles a
devil's staircase, mimicking the multi-fractal structure of the gaps
in an energy spectrum that is a Cantor set \cite{Harper}.

\begin{figure}[htbp]
\includegraphics[width =0.8\linewidth,height=0.9\linewidth]{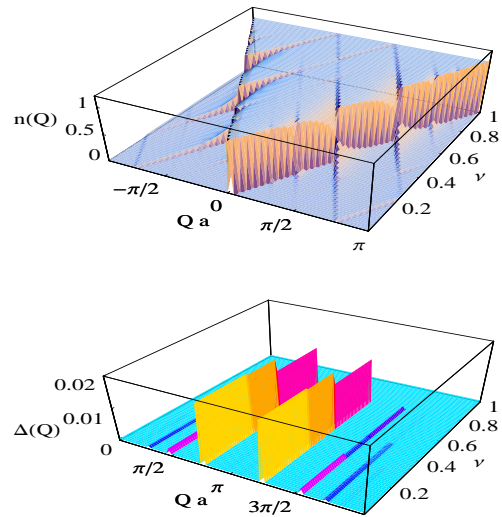}
\leavevmode \caption{ (color online) Fermions in the metallic
phase, $\lambda=0.5J$. Top: Momentum distribution and Bottom: noise
correlations.} \label{fig4}
\end{figure}

In summary, quasiperiodicity can be exploited to distinguish glassy
and Mott phases and induce a series of Mott-like band insulating phases.
Furthermore, it provides an explicit probe to witness fermionization
of bosons, and facilitates the experimental observation of complex
fractal structures.


\begin{references}
\bibitem{disorder} R. Scalettar, G. Batrouni and G. Zimanyi, Phys Rev Letter, \textbf{66}, 3144, (1991).
\bibitem{Lye} J. E. Lye {\it et al}, Phys. Rev. Lett. \textbf{95}, 070401 (2005)
\bibitem{Drese}K. Drese and M. Holthaus, Phys Rev Letter, \textbf{78}, 2932, (1997).
\bibitem{Zoller}D. Jaksch and P. Zoller, New. J. Phys. \textbf{5}, 56 (2003);
\bibitem{QPnew} B. Damski et all, Phys Rev Letter, \textbf{91}, 080403 (2003);
K. Osterloh et al, Phys Rev Letter, \textbf{95}, 010403(2005).
\bibitem{GR} M. Girardeau, J. Math. Phys. \textbf{1}, 516 (1960).
\bibitem{Paredes} B. Paredes{\it et al} Nature \textbf{429}, 277 (2004).
\bibitem{Fesh} S. L. Cornish {\it{et al}}, Phys. Rev. Lett. \textbf{85}, 1795 (2000).

\bibitem{Harper} For a review, see J. B. Sokoloff, Phys. Rep. \textbf{126}, 189 (1985).
\bibitem{HBT}R. Hanbury Brown, R. Q. Twiss, Nature \textbf{177}, 27 (1956).
\bibitem{Altman} E. Altman {\it{et al}}, Phys. Rev. A \textbf{70}, 013603 (2004).
\bibitem{Foelling} S. Foelling {\it{et al}}, Nature \textbf{434},481 (2005).
\bibitem{reynoise}A. M. Rey, I.I. Satija, and C. W. Clark {\it cond-mat}/0511700 (J. Phys. B in press).
\bibitem{Fisher} M. Fisher \textit{et al}, Phys. Rev. B \textbf{40}, 546, (1989).
\bibitem{Ott} E. Ott, {\it Chaos in Dynamical Systems}, Cambridge University Press, 1993.
\bibitem{Roth}R. Roth and K. Burnett, Phys Rev A, \textbf{68}, 023604 (2003).
\bibitem{LM}E. Lieb, T. Schultz and D. Mattis, Ann. Phys. \textbf{16} ( 1961) 407.
\bibitem{Friedel} J. Friedel,  Nuovo Cimento Suppl. \textbf{7} 287
1958.
\end{references}
\end{document}